\documentclass{article}
\newfont{\my}{msbm10 at 10pt} 
\newcommand{\mfrac}[2]
{\mbox{\footnotesize$\displaystyle\frac{\raisebox{-0.1em}{\mbox{$#1$}}}{#2}$}}
\newcommand{\wpA}{{\wp}_{\!\!{}_A}}
\newcommand{\wpO}{\wp_{\!\raisebox{-0.4ex}{\mbox{\tiny$\Omega$}}}}
\newcommand{\wpalpha}{\wp_{\!\raisebox{-0.3ex}{\mbox{\scriptsize$\alpha$}}}}

\newcommand{\wppO}{{\wp'}_{\!\!\!{}_\Omega}}
\newcommand{\wppalpha}{{\wp'}_{\!\!\!\raisebox{-0.1ex}
                                      {\mbox{\scriptsize$\alpha$}}}}
\title{Elliptic Solitons with Free Constants\\and their 
Isospectral Deformations}
\author{Yurii V$\!$.\,Brezhnev\\Theoretical Physics Department,\\
Kaliningrad State University,\\A.\,Nevsky st.\,14, 236041, Kaliningrad, 
Russia\\brezhnev@mail.ru\\[2ex]
\thanks{Supported by Russian Basic Research Foundation with Grant 00--01--00782}}

\begin{document}
\maketitle
\begin{abstract}
New exact solvable elliptic potentials with free constants 
for the spectral problems
of the third order are found. 
A time dependence of such potentials gives their isospectral
deformations and solutions of nonlinear integrable equations.
\end{abstract}


\section{Introduction}
\noindent 
In 1987, Verdier \cite{treb1, treb2} found new finite-gap potential
for the Schr\"odinger equation
\begin{equation}
\Psi''-u(x)\,\Psi=\lambda\,\Psi
\end{equation}
in elliptic functions $u(x)=6\,\wp(x)+2\,\wp(x-\omega_i)$,
differed from the well known Lam\'e potentials $u=n(n+1)\,\wp(x)$  and their
isospectral deformations. Soon, the wide classes of new elliptic solitons
(in terminology of Treibich and Verdier) were found. Later,
three main approachs in the theory of elliptic solitons have been developed,
including  the scalar and matrix spectral problems.
1) The classical approach of Hermite--Halphen, based on the  ansatz for the 
$\Psi$-function [3--7]; 
2) An analysis of the algebraic curves of the special form ---
curves being a covering over tori \cite{smirnov1, enol3, monograph};
3) The methods, touching the Picard's theorem and hierarchies of the higher
stationary nonlinear differential equations [9--13]. 

There are the following main problems in the theory of elliptic solitons:
 the search for the potentials $u_i(x)$, formula for the $\Psi$-function,
corresponding algebraic curve and its realization in the form of 
covering of torus, isospectral deformations of $u_i$ with respect to
integrable nonlinear equations.

Originally (1872), Hermite developed an approach to the integration of the 
Lam\'e potentials for (1) using an  ansatz for the $\Psi$-function
as a meromorphic two-periodic function of the  second kind by Picard 
\cite{whitt, akhiezer} (Hermite used Jacobi's 
$H,\,\Theta, \,Z$-functions):
\begin{equation}\label{ans1}
\Psi=\frac{\sigma(x-\alpha_1^{}) \cdots \sigma(x-\alpha_m^{})}
{\sigma(x-\Omega_1^{}) \cdots \sigma(x-\Omega_m^{})}\,e^{k(\lambda) x},
\end{equation}
where the poles $\Omega_j$ of the potential $u_i(x)$ independent of $\lambda$.
In such ansatz, zeroes of the $\Psi$-function $\alpha_i^{}$ depend on the 
spectral parameter $\lambda$ and are determined from  the {\em several\/} 
essentially 
{\em transcendental\/} equations. By this reason the ansatz (\ref{ans1})
is not convenient and not effective. Later, Hermite and Halphen considered the following
ansatz
\begin{equation}\label{ans2}
\Psi=e^{kx}\sum_{n,\,j}\, a_{nj}\,\Phi^{(n)}(x-\Omega_j;\alpha),
\end{equation}
where the $\Phi$-function ({\em \'el\'ement simple} by Halphen)
is defined by the expression
$$
\Phi(x;\alpha)=\frac{\sigma(\alpha-x)}{\sigma(\alpha)\,\sigma(x)}\,
e^{\zeta(\alpha) x}.
$$
Thus, instead of unknown parameters $\alpha_i(\lambda),\,k(\lambda)$ in 
(\ref{ans1})
we have  only two ones $k,\, \alpha$ in the framework of ansatz (\ref{ans2}).
Indeed, after the substitution (\ref{ans2}) into the spectral problem
\begin{equation}\label{L}
\hat L \Psi \equiv \frac{d^n}{dx^n}\Psi(x;\lambda)+u_1(x,\lambda)\,
\frac{d^{n-1}}
{dx^{n-1}}\Psi(x;\lambda)
+\cdots+u_n(x,\lambda)\,\Psi(x;\lambda)=0
\end{equation}
and expanding the principal part of poles at $x=\Omega_j$, we will 
have  equations being polynomial in $k$ and transcendental  in $\alpha$.
In fact, $\alpha$ is a parameter of the covered torus
$$
\wp'(\alpha)^2=4\,\wp(\alpha)-g_2^{}\,\wp(\alpha)-g_3^{}.
$$
Evidently, that the obtained equations are linear ones with respect to
$a_{nj}$ and therefore  elementary solvable.

The classical potentials of Lam\'e, Halphen, Treibich--Verdier and others
have a fixed location of poles. So, considering  the potential 
$$
u(x)=2\,\wp(x)+2\,\wp(x-a)+2\,\wp(x-b)
$$
for the spectral problem (1), we get the restriction: $b=-a, \, \wp''(a)=0$.
In this paper, we find  new potentials containing free parameters 
and  poles (see also the most recent information in \cite{treb3}).
Assuming their dependence on time $t$ we will obtain the isospectral deformation
of initial condition $u(x)$ as a solution of the corresponding nonlinear
integrable equation.

\section{Algorithm for the solution of the spectral problems}
\noindent 
The general algorithm for integrating  the spectral problem (\ref{L})
with elliptic coefficients $u_i$ of $x$ and arbitrary  of $\lambda$ 
was proposed
by the author in \cite{brezhnev}. It has a pure algebraic characterization and
is reduced to a system of polynomial equations solvable by the polynomial 
algebra
methods (Gr\"obner bases, characteristic sets and others \cite{buch, coh}).
Indeed, the equations on  poles of the expression $\hat L \Psi$,
where the $\Psi$-function is given by the formula (\ref{ans2}), contain
the following unknown quantities:
$$
k, \, \wp(\alpha), \, \wp'(\alpha), \, \wp(\Omega_i-\Omega_j), \, 
\wp'(\Omega_i-\Omega_j), \, 
 \zeta(\alpha), \, 
\zeta(\Omega_i-\Omega_j), \, \Phi(\Omega_i-\Omega_j; \alpha).
$$
Using the addition theorems for elliptic functions and two important 
 equalities
$$
\Phi(x+z; \alpha)=\frac12\,\frac{\Phi(x; \alpha)\,\Phi(z; \alpha)}
{\wp_{\!\raisebox{-0.3ex}{\mbox{\scriptsize$x$}}}-
\wp_{\!\raisebox{-0.3ex}{\mbox{\scriptsize$z$}}}}
\left(
\frac {\wppalpha+
{\wp'}_{\!\!\!\raisebox{-0.1ex}{\mbox{\scriptsize$x$}}}}{\wpalpha-
\wp_{\!\raisebox{-0.3ex}{\mbox{\scriptsize$x$}}}}-
\frac {\wppalpha+
{\wp'}_{\!\!\!\raisebox{-0.1ex}{\mbox{\scriptsize$z$}}}}{\wpalpha-
\wp_{\!\raisebox{-0.3ex}{\mbox{\scriptsize$z$}}}}
\right)\!,
\quad
\Phi(\Omega;\alpha)\,\Phi(-\Omega;\alpha)=\wpalpha-\wpO,
$$
where we introduce the short notation $\wpalpha  \equiv \wp(\alpha)$ etc.,
the above mentioned equations on poles are ``algebraized'': they are 
transformed into the polynomial ones with respect to the variables 
$k, \, \wp(\alpha), \, \wp'(\alpha), \, \wp(\Omega_i), \, \wp'(\Omega_i)$
(see \cite{brezhnev} for the details).
Thus, we have a polynomial ideal in a  ring 
$\mbox{\my Q}(\lambda, g_2^{}, g_3^{})[k,\wpalpha, \wppalpha, \wpO, 
\wppO]$ and the final solution of the spectral problem (\ref{L}) 
is defined by the structure of this ideal \cite{progr, coh, brezhnev}.

\section{Elliptic solitons with free constants}
\noindent 
Let us consider the spectral problem
\begin{equation}\label{sk}
\Psi''' - u(x)\, \Psi'=\lambda\,\Psi.
\end{equation}
Here, the  potentials with only pole in a parallelogramm of periods have not 
unified form like the Lam\'e potentials. Indeed, the general
1-pole elliptic potential has the form
\begin{equation}\label{1pole}
u(x)=A\,\wp(x)+B.
\end{equation}
Investigating (\ref{1pole})  in the framework of the ansatzs
\begin{equation}\label{ans3}
\Psi=\left(\Phi(x;\alpha)+a_1^{}\,\Phi'(x;\alpha)+\cdots+
a_n^{}\,\Phi^{(n)}(x;\alpha)\right) e^{kx}
\end{equation}
under $n=0,1,2,3,4$ we get: 
\begin{itemize}
\item $n=0$: \ $u=6\,\wp(x)+B$ \cite[p.\,509]{kamke}. 
$B$ is an arbitrary constant, genus $g=2$;

\item $n=1$: \ $u=12\,\wp(x)$. The degenerated curve is of genus $g=1$;

\item $n=2$: \ No solution. $\lambda$ must be zero \cite{enol2};

\item $n=3$: \ $u=30\,\wp(x) \pm 3\sqrt{\mathstrut 3\,g_2^{}}$. Genus $g=3$;

\item $n=4$: \ $u=42\,\wp(x)$. Genus $g=6$.
\end{itemize}
All the associated curves are the trigonal ones. 
$\lambda$ is a meromorphic function with 
only 3-rd order pole. 
The 2-pole elliptic potentials in (\ref{sk}) have the form
$$
u=A\,\wp(x)+B\,\wp(x-\Omega)+C\,\zeta(x)-C\,\zeta(x-\Omega)+D.
$$
Using the $\Theta$-functional approach, one can show that the finite-gap potential
for (\ref{sk}) is a second logarithmic derivative of $\Theta$ and thus, 
must not has residues. Therefore $C=0$.
The simplest nontrivial example is
\begin{equation}\label{2pole}
u=6\,\wp(x)+6\,\wp(x-\Omega).
\end{equation}
The corresponding solution is as follows \cite{brezhnev}:
\begin{equation}\label{fullnew}
\begin{array}{c}
\Psi(x)=\Big(a\,\Phi(x;2\,\alpha)+\Phi(\Omega;2\,\alpha)\,
\Phi(x-\Omega;2\,\alpha)\!\Big) e^{kx},\\
a=
\displaystyle
\zeta(2\,\alpha+\Omega)- 2\,\zeta(\alpha)-\zeta(\Omega)^
{\displaystyle {}^{\mathstrut}},
\quad 
k=2\,\zeta(\alpha)-\zeta(2\,\alpha),\quad \lambda=-4\,\wp'(\alpha).
\end{array}
\end{equation}
The generalization of Halphen's equation
$$
\Psi'''-((n^2-1)\,\wp(x)+h)\,\Psi'-\mfrac{n^2-1}{2}\,\wp'(x)\,\Psi=
\lambda\,\Psi
$$
was considered by Unterkofler \cite{unterkofler} in the framework of 
the ansatz
(\ref{ans1}). The algebraic curves were written, but  important 
formulas for the 
covering of torus
$R(\wp(\alpha), \wp'(\alpha), \lambda)=0$ (i.e. the transcendental equation 
for the determining of $\alpha$ by $\lambda$) have not been obtained.

Let us consider the spectral problem
\begin{equation}\label{uv}
\Psi'''-u(x)\,\Psi'-v(x) \Psi=\lambda\,\Psi.
\end{equation}
The most general 1-pole elliptic potential for the equation (\ref{uv}) 
has the form:
\begin{equation}\label{strange}
\Psi'''-( a\,\wp(x)+d)\,\Psi'-
(b\,\wp'(x)+c\,\wp(x))\,\Psi=\lambda\,\Psi.
\end{equation}
The case $n=0$ in the ansatz (\ref{ans3}) yields the solution:
$$
\Psi'''-3\,(\wp(x)+c^2)\,\Psi'-
\left(\mfrac{3}{2}\,\wp'(x)+3\,c\,\wp(x)\right) \Psi=\lambda\,\Psi,
\qquad 
\Psi=\Phi(x;\alpha)\,e^{cx}.
$$
Under $n=1$ in (\ref{ans3}) we arrive at the nontrivial solution. 
Indeed, the first of three polynomials in the original base has the form:
\begin{equation}\label{A12}
(a-12)\,\Big(\!(a-18)\,(k^2-\wpalpha) + 2\,c\,k\Big)+ 2\,(a-18)\,d+c^2=0,
\end{equation}
where without  loss of generality we set $b=12-a$. In general case, 
an attempt to solve 
the corresponding system of algebraic equations   failed.
For example, in some particular cases, all the solutions lead to the 
dependence of parameters
$a,\,c,\,d$ on spectral one $\lambda$. But under $a=12$ the polynomial 
(\ref{A12})
is simplified: the dependence on $k,\,\alpha$ disappeares. 
After the replacement $c\rightarrow 12\,c$, the  determining
system of equations takes the form
\begin{equation} \label{base}
\small 
\!\!\!\!\!\!\! 
\begin{array}{l}
\displaystyle 3\,k^4-4\,c \,k^3-6\, (3\,\wp+2\,c^2 )\,k^2+ 
12\,(c\,\wp+\wp' )\,k-
9\, \wp^2 + 12\,c^2\, \wp - 4\,c\, \wp' - 2\,\lambda\,c  + 3\,g_2^{}=0,\\
\displaystyle 4\,k^{3^{\mathstrut}}-6\,c\,k^2-12\, (\wp+2\,c^2 )\,k+6\,c\,
\wp+4\,\wp'+24\,c^3-\lambda=0,\\
\wp'^{2^{\mathstrut}}-4\,\wp^3+g_2^{}\,\wp+g_3^{}=0
\end{array}
\end{equation}
and the solution for the remaining variables read as
\begin{equation} \label{c}
\Psi'''-12\,(\wp(x)+c^2)\,\Psi'- 12\,c\,\wp(x)\,\Psi = \lambda\,\Psi,
\quad 
\Psi=\Big(\!(k-2\,c)\,\Phi(x;\alpha)+\Phi'(x;\alpha)\!\Big) e^{kx}.
\end{equation}
The ideal (\ref{base}) is not equal $ \langle 1\rangle$ in a ring 
$\mbox{\my Q}(\lambda,c,g_2^{},g_3^{})[\wp, \wp'\!, k]$. Thus, $c$ is an 
arbitrary constant. For a completeness we give the expression for the 
algebraic curve $F(k,\lambda)=0$ depended on $c$:
\begin{equation}\label{bigcurve}
\small           
\!\!\!\!\!\!\begin{array}{l}
\phantom{\frac{}{{}_{\mathstrut}}}
64\, (\lambda-11\,c^3 ) (\lambda^2+32\,\lambda\,c^3+ 
2^8\,c^6-108\,g_2^{}\, c^2 ) \underline{\,k^3_{}\,}-48\, (c\, \lambda^3+
4\, (3\,c^4+g_2^{} )\,\lambda^2-\\
\phantom{\frac{}{{}_{\mathstrut}}}
4\, (96\,c^{6}+13\,g_2^{}\,c^2-36\,g_3^{}
 )\,c\,\lambda-16\, (320\,c^8-127\,g_2^{}\,c^4+99\,
g_3^{}\,c^{2}+9\,g_2^2 )\,c^2 ) \underline{\,k^2_{}\,}-\\
\phantom{\frac{}{{}_{\mathstrut}}}
48\, (11\,c\,\lambda^3+4\, (60\,c^4+g_2^{} )\,\lambda^2-
4\, (192\,c^6+301\,g_2^{}\,c^2-72\,g_3^{} )\,c\,\lambda-\\
\phantom{\frac{}{{}_{\mathstrut}}}
16\, (1792\,c^8-676\,g_2^{}\,c^4+198\,g_3^{}\,c^2-9
\,g_2^2 )\,c^2 )\,c \underline{\,k_{}\,} -27\,\lambda^4-352\,c^3\,\lambda^3+\\
\phantom{\frac{}{{}_{\mathstrut}}}
24\, (160\,c^6+151\,g_2^{}\,c^2-36\,
g_3^{})\,\lambda^2-384\, (128\,c^8+112\,g_2^{}\,c^4
-18\,g_3^{}\,c^2+3\,g_2^2 )\,c\,\lambda-\\
\phantom{\frac{}{{}_{\mathstrut}}}
2^{17}11\,c^{12}+
2^{10}609\,g_2^{}\,c^8-2^8891\,g_3^{}\,c^6-25776\,g_2^2\,c^4+
2^7 3^5\,g_2^{}\,g_3^{}\,c^2+2^8\,(g_2^3-27\,g_3^2)=0.
\end{array}
\end{equation}
This curve has 8 finite points of ramification $\lambda_i$ with indexes
(2,1) and a ramification at infinity with index 3. Genus $g=3$.
The curve is not hyperelliptic one. The corresponding covering of torus
$R(\wp(\alpha), \wp'(\alpha), \lambda)=0$ is as follows:
\begin{equation}\label{bigcovering}
\small 
\begin{array}{l}
\phantom{\frac{}{{}_{\mathstrut}}}
32\, (\lambda-11\,c^3 ) (\lambda^2+32\,
\lambda\,c^3+ 2^8\,c^6-108\,g_2^{}\,c^2 ) \underline{\,\wp'(\alpha)\,}-
48\, (c\,\lambda^3+4\, (3\,c^4+g_2^{})\,\lambda^2-\\
\phantom{\frac{}{{}_{\mathstrut}}}
4\, (96\,c^6+13\,g_2^{}\,c^2-36
\,g_3^{} )\,c\,\lambda-16\, (320\,c^8-127\,g_2^{}\,c^4+
99\,g_3^{}\,c^2+9\,g_2^2 )\,c^2 ) \underline{\,\wp(\alpha)\,} + \\
\phantom{\frac{}{{}_{\mathstrut}}}
\lambda^4+32\,c^3\,\lambda^3-24\, (g_2^{}\,c^2+12\,g_3^{} )\,
\lambda^2-128\, (64\,c^8+6\,g_2^{}\,c^4-9\,g_3^{}\,c^2+3\,g_2^2 )\,
c\,\lambda-\\
\phantom{\frac{}{{}_{\mathstrut}}}
2^{16}\,c^{12}-2^{11}3\,g_2^{}\,c^8-94464\,g_3^{}\,c^6+5520\,g_2^2\,c^4+
2^7 81\,g_2^{}\,g_3^{}\,c^2+2^8\,(g_2^3-27\,g_3^2)=0.
\end{array}
\end{equation}
This equation as an elliptic function 
of $\alpha$ of the 3-rd order has three solutions $\alpha_{{}_{1,2,3}}$ corresponding to
three linear independent solutions of (\ref{c}) with arbitrary $\lambda,\,c$.
To all appearances, the potential (\ref{c}) corresponds to the higher  Boussinesq
equations, but  its complete finite-gap treatment is  an open question.
We revealed one more new case, when a polynomial system is simplified
and a solution is not trivial. It is
$$
\Psi'''-(18\,\wp(x)+d)\,\Psi' +6\,\wp(x)\,\Psi = \lambda\,\Psi,
\quad 
\Psi=\Big(\!(3\,k^2+3\,\wpalpha-d)\,\Phi(x;\alpha)+6\,k\,\Phi'(x;\alpha)
\!\Big) e^{kx}.
$$
The corresponding formulas for curve and covering type as 
(\ref{bigcurve}) and  (\ref{bigcovering}) of genus $g=4$ are too big that 
to display here. Another case 
\begin{equation}\label{tmp}
\psi'''-(6\,\wp(x)+d)\,\psi' - 6\,\wp'(x)\,\psi = \lambda\,\psi
\end{equation}
is trivial in the sense that $\psi=\Psi'$ in (\ref{sk}) with $n=0$ 
is the solution of (\ref{tmp}).

The next nontrivial example is a simplest 2-pole potential in (\ref{uv}).
\begin{equation}\label{33}
\Psi'''-3\,(\wp(x)+\wp(x-\Omega)-\wpA)\,\Psi'-
\left. \left.\mfrac32 \right(\! \wp'(x)+\wp'(x-\Omega)+
B\,\wp(x)- B\,\wp(x-\Omega)\right) \Psi=\lambda\,\Psi.
\end{equation}
This equation is integrable with arbitrary $\Omega$ and $A$ 
(see \cite{brezhnev, smirnov2} for the solution) and the restriction
$B^2=4\,\wpO - 4\,\wpA$.

\section{Isospectral deformations}
\noindent 
If $\Psi$-function depends on a parameter $t$, then
the arbitrariness of location of poles $\Omega_i$ in potential
means its isospectral deformation with $\Omega_i(t)$. Let us consider
 examples.

The spectral problem (\ref{sk}) corresponds to the Sawada--Kotera equation
\cite{sawada}
\begin{equation}\label{sawada}
u_t^{}=u_{\mathit{xxxxx}} - 5\,u\,u_{\mathit{xxx}} 
-5\,u_x\, u_{\mathit{xx}}+5\,u^2\,u_x.
\end{equation}
A deformation of the potential (\ref{2pole}) is stationary.
Chazy \cite[p.\,380]{chazy} found the more general stationary solution of 
the equation (\ref{sawada})
$u(x,t)=6\,\wp_1^{}+6\,\wp_2^{}$,
where we introduced the designations:
$$
\wp_1^{} \equiv \wp(x-c\,t-\Omega;\,g_2^{},g_3^{}), \quad
\wp_2^{} \equiv \wp(x-c\,t-\tilde\Omega;\,g_2^{},\tilde g_3^{}), \quad
c=-12\,g_2^{}.
$$
Now we use a connection between the equation (\ref{sawada}) and 
the Kaup--Kupershmidt, Fordy--Gibbons \cite{fg} and  Tzitzeica equations
respectively
\begin{equation}\label{kk}
w_t=w_{\mbox{\scriptsize\it xxxxx}} - 5\,w\,w_{\mbox{\scriptsize\it xxx}} 
- \mfrac{25}{2}\,w_x w_{\mbox{\scriptsize\it xx}}
+ 5\,w^2\,w_x,
\end{equation}
\begin{equation}\label{fg}
v_t=v_{\mbox{\scriptsize\it xxxxx}}-5\,(
v_x\,v_{\mbox{\scriptsize\it xxx}}+
v^2\,v_{\mbox{\scriptsize\it xxx}}+
v_{\mbox{\scriptsize\it xx}}^2+
v_{\mbox{\scriptsize\it x}}^3+
4\,v\,v_x\,v_{\mbox{\scriptsize\it xx}}-
v^4\,v_x),
\end{equation}
\begin{equation}\label{tz}
\phi_{\mbox{\scriptsize\it xt}}=e^\phi-e^{-2\,\phi\displaystyle \mathstrut}
\end{equation}
by means of the Miura maps
\begin{equation}\label{miura}
u=v^2-v_x,\qquad w=v^2+2\,v_x, \qquad u=\phi_{\mbox{\scriptsize\it xx}}+
\phi_x^2.
\end{equation}
After the substitution $v=-\ln'\!\psi$,
the first Miura's map in (\ref{miura}) is linearized and we arrive at 
a problem
of integration of the Schr\"odinger equation without the spectral parameter:
$$
\psi_{\mathit{xx}}=6\,(\wp_1^{}+\wp_2^{})\,\psi.
$$
Using the solution (\ref{fullnew}) under $\lambda=0$ (the passage to the limit
$\alpha \rightarrow \omega_i$) we will have
$$
\Psi= C_1^{}\,\Big(\zeta(x)-\zeta(x-\Omega)\!\Big) +C_2^{}.
$$
After  the substitution
$\Psi'=\psi$, we get the result 
(under the  coinciding invariants $g_3^{}=\tilde g_3^{}$):
$$
\psi=\wp_1^{}-\wp_2^{}.
$$
The straightforward verification shows that it will be also valid under 
$g_3^{} \ne \tilde g_3^{}$. Thus we obtain the 5-parametric stationary 
solutions of the equations (\ref{kk}, \ref{fg}):
$$
v(x,t)=\frac{{\wp'}_{\!\!1}-{\wp'}_{\!\!2}}{\wp_1^{}-\wp_2^{}},
\qquad
w(x,t)=-12\,(\wp_1^{}+\wp_2^{})+ 3\left( 
\frac{{\wp'}_{\!\!1}-{\wp'}_{\!\!2}}{\wp_1^{}-\wp_2^{}}
\right)^2.
$$
The same technique leads to the solution of the Tzitzeica 
equation (\ref{tz}), but the 
time dynamics is not stationary:
$$
\phi(x,t)=\ln 2\,c+
\ln\!\Big(\wp(x+c\,t-\Omega; \,g_2^{}, g_3^{}) - 
\wp(x-c\,t-\tilde \Omega; \,g_2^{}, \tilde g_3^{})\!\Big),
\qquad 4\,(\tilde g_3^{}-g_3^{})\,c^3=1.
$$

Let us consider the time deformation of 2-pole potential (\ref{33}) 
\begin{equation}\label{deform}
u(x,t)=3\,\wp(x-\Omega_1^{})+3\,\wp(x-\Omega_2^{})-3\, \wpA
\end{equation}
with respect to the Boussinesq equation
\begin{equation}\label{bouss}
3\,u_{\mathit{tt}}^{}=\left(2\,u^2-u_{\mathit{xx}}\right)_{\mathit{xx}}.
\end{equation}
Substituting the ansatz (\ref{deform}) in (\ref{bouss}) and equating the
poles to zero we will obtain the equations on the poles $\Omega_{1,2}^{}(t)$ 
--- the Calogero--Moser system of two particles with a repulsion potential and 
immovable center of mass:
\begin{equation}\label{CM}
\left\{
\begin{array}{ccc}
\ddot\Omega_{1_{\displaystyle\mathstrut}}-4\,\wp'(\Omega_1^{}-\Omega_2^{})&=&0\\
\ddot\Omega_2^{}+4\,\wp'(\Omega_1^{}-\Omega_2^{})&=&\phantom{,}0,
\end{array}
\right.
\qquad
\left\{
\begin{array}{ccc}
\dot\Omega_1^2-4\,\wp(\Omega_1^{}-\Omega_2^{})+4\,\wpA &=&0\\
\dot\Omega_2^{2^{\displaystyle\mathstrut}}-4\,\wp(\Omega_1^{}-\Omega_2^{})+4\, 
\wpA &=& \phantom{,}0.\\
\end{array}
\right.
\end{equation}
Ruling out its trivial solution $\Omega_1^{}-\Omega_2^{}=C$, 
we get $\Omega_2^{}=C-\Omega_1^{}$ and an equation on $\Omega_1$:
\begin{equation}\label{uni1}
\dot\Omega_1^2=4\,\wp(2\,\Omega_1^{}-C)-4\,\wpA.
\end{equation}
The substitution $X=\wp(2\,\Omega_1^{}-C)$ transforms the equation (\ref{uni1})
into the algebraic form:
\begin{equation}\label{uni2}
\dot X^2 = 16\,(4\,X^3-g_2^{}\,X-g_3^{})(X-\wpA).
\end{equation}
(\ref{uni2}) is the algebraic curve $F(X,\dot X)=0$ of genus 1. 
After the birational change $X \to \tilde X $
$$
X={\frac {3\,\wpA\,\tilde X-8\,g_2^{}\,\wpA-12\,g_3^{}}
{3\,\tilde X- 48\,\wpA^2+4\,g_2^{}}}
$$
we obtain an equation on $\tilde X$ in the canonical Weierstrass's form
$$
\left(\!\frac{d\tilde X}{d t}\!\right)^2=
4\, \tilde X^3-\tilde g_2^{}\,\tilde X-\tilde g_3^{},\qquad
\tilde X(t) = \wp(t-t_0^{}; \,
\tilde g_2^{},\, \tilde g_3^{})  \equiv \tilde \wp(t-t_0^{})
$$
with new invariants
$$
\tilde g_2^{} = 2^8\,g_2^{}\,\wpA^2+768\,g_3^{}\,\wpA+
{\mfrac {64}{3}}\,g_2^2,
$$
$$
\tilde g_3^{} = -2^{12}\,g_3^{}\,\wpA^3-{\mfrac {2^{11}}{3}}\,
g_2^2\,\wpA^2-2^{10}\,g_2^{}\,g_3^{}\,\wpA+
\mfrac {2^9}{27}\,g_2^3-2^{10}\,g_3^2.
$$
Thus, we get the complete solution of the equations (\ref{CM}):
$$
\Omega_1^{}(t)=\mfrac12\,\wp^{-1}\! \!\left(
{\frac {3\,\wpA\,\tilde \wp(t-t_0^{})-8\,g_2^{}\,\wpA-12\,g_3^{}}
{3\,\tilde \wp(t-t_0^{})- 48\,\wpA^2+4\,g_2^{}}}
\right)+\mfrac C2.
$$
$t_0^{},\,C$ are arbitrary constants and 
the elliptic integral $\wp^{-1}$
corresponds to the invariants $g_2^{}, \,g_3^{}$.
See also \cite{chud} for the solution in the context of the solutions
of the Kadomtsev--Petviashvili equation.

\end{document}